\documentclass[aps,prm,reprint,groupedaddress,longbibliography]{revtex4-2}
\usepackage{mathptmx,graphicx,amsmath,pifont,epstopdf,amssymb}
\usepackage[version=4]{mhchem}
\usepackage[T1]{fontenc}
\usepackage[colorlinks,linkcolor=blue, urlcolor=blue, anchorcolor=blue, citecolor=blue]{hyperref}
\usepackage{xcolor}  
\colorlet{RED}{red}
\makeatother
\begin{document}
\title{Generalized Phase Diagrams for Graphene CVD growth on Copper}
\author{Tongtong Wang}
\affiliation{National Laboratory of Solid State Microstructures and School of Physics, Nanjing University, Nanjing 210093, China}
\affiliation{Collaborative Innovation Center of Advanced Microstructures, Nanjing University, Nanjing 210093, China}
\author{Ke Jin}
\affiliation{National Laboratory of Solid State Microstructures and School of Physics, Nanjing University, Nanjing 210093, China}
\affiliation{Collaborative Innovation Center of Advanced Microstructures, Nanjing University, Nanjing 210093, China}
\author{Yishi Zhang}
\affiliation{National Laboratory of Solid State Microstructures and School of Physics, Nanjing University, Nanjing 210093, China}
\affiliation{Collaborative Innovation Center of Advanced Microstructures, Nanjing University, Nanjing 210093, China}
\author{Dajun Shu}
\affiliation{National Laboratory of Solid State Microstructures and School of Physics, Nanjing University, Nanjing 210093, China}
\affiliation{Collaborative Innovation Center of Advanced Microstructures, Nanjing University, Nanjing 210093, China}
\date{\today}

\begin{abstract}

Understanding the competition between first-layer lateral expansion and second-layer nucleation is essential for layer-controlled graphene growth via chemical vapor deposition (CVD). Building on our previous phase diagram framework based on the dimensionless parameters $\alpha$ and $\Gamma$, we develop an enhanced model incorporating two previously neglected effects: thermal-expansion-induced substrate strain and chemical desorption of carbon monomers via reverse dehydrogenation. First-principles calculations are employed to determine the strain-dependent diffusion and attachment barriers on both exposed and graphene-covered Cu(111) surfaces. By mapping the multi-step CVD process into an effective quasi-physical vapor deposition, we construct a generalized phase diagram characterized by the coupled effects of $\alpha$, $\Gamma$, and a newly introduced desorption parameter $Z$.
Our results show that tensile strain expands the bilayer graphene (BLG) growth window for critical nucleus sizes $i^*>1$. In contrast, chemical desorption suppresses BLG formation in the high-$\Gamma$ regime via $Z$-dependent monomer depletion. This unified framework provides a predictive guide for the rational synthesis of high-quality bilayer graphene by linking macroscopic growth parameters to microscopic layer-selection mechanisms.

\end{abstract}

\maketitle

\section{Introduction}
\label{S1}

Graphene has been a focus of materials research since its first exfoliation, owing to its extraordinary electronic, optical, mechanical, and thermal properties\cite{Novoselov-2004-Science, KimKeunSoo-2009-Nature, ShenJianhua-2012-ChemCommun, Novoselov-2007-Science, Alexander-2011-NatMater, LeeChanggu-2008-Science}. Beyond single-layer graphene (SLG), bilayer graphene (BLG) has attracted particular attention. Specifically, AB-stacked BLG exhibits a tunable band gap in a vertical electric field, while twisted BLG hosts novel correlated states such as Mott insulating state and superconductivity\cite{XiaFengnian-2010-NanoLetter, Eduardo-2007-PRL, Island-2019-Nature, Caoyuan-2018-Nature-mott, Caoyuan-2018-Nature-sc}. Chemical vapor deposition (CVD) on copper substrate has emerged as the most viable approach for the scalable production of high-quality graphene\cite{LeeHCheun-2016-ProcediaChem, Novoselov-Nature-2012}. However, achieving precise control over the layer number remains a significant challenge\cite{YaoWenqian-AFM-2022, LiuBing-Nanoscale-2024}. The unintended formation of adlayers can significantly reduce carrier mobility, thereby hindering large-scale device integration\cite{Yagi-SciRep-2018, Nagashio-APE-2009}. 

The CVD growth of graphene on catalytic surfaces is a complex multi-stage process, encompassing the decomposition of gas-phase precursors (typically $\ce{CH4}$), surface diffusion of carbon monomers, nucleation, and subsequent island growth. Extensive theoretical investigations have elucidated many of these aspects, with particular focus on precursor dissociation and surface kinetics. For $\ce{CH4}$ decomposition on Cu substrates, the catalytic pathways and the dominant growth species have been characterized in detail through first-principles calculations\cite{Grzegorz-2011-JCP, HeYingyou-2019-CMS, LiKai-2014-Carbon, ZhangWenhua-2011-JPCC, WangXinlan-2017-Nanoscale}. Regarding growth kinetics, multiscale modeling has highlighted the critical role of surface kinetics in governing the growth rate and morphology\cite{MaTeng-PNAS-2013, QiuZongyang-ACR-2018, LiPai-JPCC-2020, ChengTing-APLM-2021, LiPai-JPCC-2017, ChenShuai-Carbon-2019, KongXiao-npjCM-2021, ZhangDi-NC-2024}. In the context of BLG, the self-limiting mechanism leads to a characteristic "inverse wedding cake" growth mode, where the competition between first-layer expansion and second-layer nucleation becomes the decisive factor\cite{WeiChen-2015-PRB, HongHyoChan-Nanomaterials-2023, XueRuiwen-FML-2017, SunLuzhao-NC-2021, LiQiongyu-NanoLett-2013, NieShu-NJP-2012}. 

In our previous work, we developed a rate-equation model to capture this competition by reducing complex surface kinetics into two dimensionless parameters, $\alpha$, characterizing the competition between edge-crossing and attachment, and $\Gamma$, representing the relative rate of surface diffusion with respect to the deposition flux\cite{Wangtt-PRM-2025}. Within this framework, a phase diagram was constructed to delineate the SLG and BLG growth regimes, successfully elucidating the observed effects of hydrogen pressure. 

However, two important physical effects were omitted for simplicity in our previous framework. First, temperature can indirectly influence graphene growth through thermally induced substrate strain, which modifies both surface kinetics and cluster formation energetics\cite{WangKai-HTMS-1996, Mondal-JPCA-2013, Marashdeh-JPCC-2013, DongYibo-Nanomaterial-2019}. Second, chemical desorption via reverse dehydrogenation reactions (\emph{i.e.}, reassociation of adsorbed C and H into hydrocarbon species) can deplete surface carbon monomers under high hydrogen partial pressure\cite{Vlassiouk-ACSNano-2011, SunXiucai-NanoRes-2024, Leidinger-JPCC-2023, ZhaoPei-Nanoscale-2016}. Consequently, the coupled influence of thermal strain and chemical desorption on the kinetic parameters remains unexplored. Beyond that, recent study found that high hydrogen partial pressure shifts the nucleation balance, where hydrogen stabilizes small carbon rings in the nucleation process\cite{Mitchell-Carbon-2018}. Furthermore, a "sunk growth" mode was proposed where the second layer partially embeds into the Cu surface \cite{DaiXinyue-Carbon-2022, XuZiwei-npjCM-2020, ChenBuhang-NatCommun-2025}. Although such effects are primarily relevant for thicker adlayers and do not significantly influence the present bilayer regime. A unified physical picture integrating reversible decomposition under high hydrogen pressure with temperature-dependent surface kinetics is thus essential, not only for refining growth precision at elevated temperatures but also for providing a mechanistic roadmap for low-temperature graphene synthesis.

In this work, we extend our previous rate-equation framework by incorporating strain-modulated kinetics and the chemical desorption of C monomers.
Through systematic first-principles calculations, we evaluate the kinetic barriers and cluster stabilities under biaxial strain. By deriving effective deposition and desorption rates ($F_{\text{eff}}$ and $z_{\text{eff}}$), the multi-step CVD process is transformed into an effective quasi-PVD process. 
We establishes a generalized kinetic framework for layer-selective graphene growth, in which the competition between first-layer expansion and second-layer nucleation is governed by the coupled effects of $\alpha$, $\Gamma$, and the newly introduced desorption parameter $Z$. Within this framework, we further show that tensile-strained Cu(111) favors BLG growth for critical nucleus sizes $i^*>1$. Chemical desorption further modifies the phase boundary in the high-$\Gamma$ regime through the additional kinetic suppression introduced by the parameter $Z$. These findings establish a robust theoretical foundation for the rational design of CVD parameters to synthesize high-quality single-layer or bilayer graphene.

\section{Models and Methods}
\label{S2}
Following our previous work, graphene growth on Cu(111) is described using a coarse-grained kinetic framework in which carbon monomers are treated as the dominant growth species. To distinguish the kinetic environments associated with different graphene layers, the Cu surface is divided into exposed and covered substrate regions depending on whether the surface is directly covered by graphene islands. Kinetic processes occurring on the exposed substrate are denoted as first-layer events (subscript surf), whereas processes taking place beneath the graphene island are denoted as second-layer events (subscript gap).

\begin{figure}[tb]
\centering
\includegraphics[width=\linewidth]{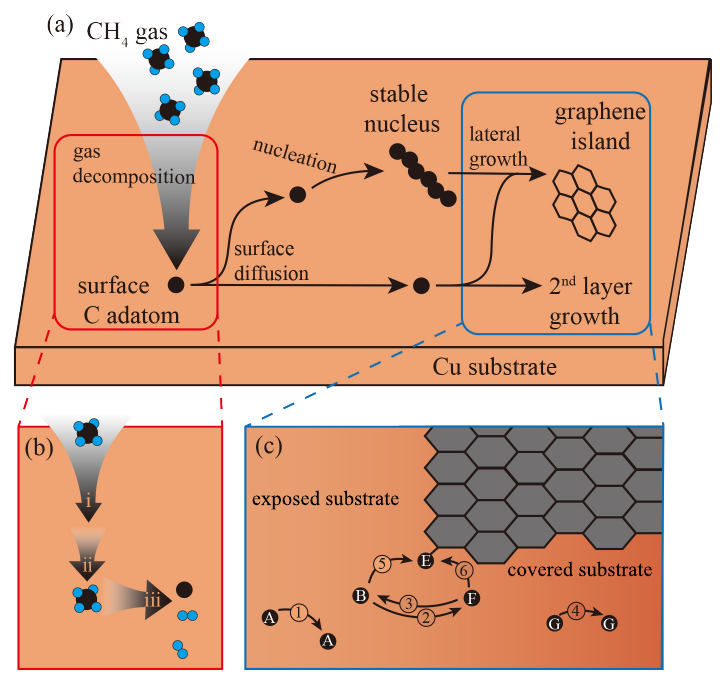}
\caption{(a) Schematic diagram of all processes involved in graphene growth. The red rectangle highlights the decomposition reaction process, while the blue rectangle highlights the surface growth process. (b) Detailed view of the gas reaction processes, which involves three steps. (c) Detailed view of the surface growth processes. Six processes are considered in this model.}
\label{F1}
\end{figure}

Figure~\ref{F1}(a) summarizes the kinetic processes involved in graphene CVD growth, which are separated into two coupled stages, namely the \textit{precursor decomposition stage} (red box) and the \textit{surface growth stage} (blue box). As illustrated in Fig.~\ref{F1}(b), the precursor decomposition stage includes (i) gas-phase transport of $\ce{CH4}$, (ii) transport across the vapor--substrate interface, and (iii) sequential dehydrogenation reactions that generate surface carbon monomers and $\ce{H2}$. The carbon monomers subsequently participate in the surface growth stage shown in Fig.~\ref{F1}(c), which includes monomer diffusion on the exposed substrate (A $\leftrightarrow$ A, process no.~1), edge crossing between exposed and covered substrate regions (B $\leftrightarrow$ F, processes no.~2 and no.~3), monomer diffusion beneath the graphene island (G $\leftrightarrow$ G, process no.~4), and monomer attachment to the graphene edge from either the exposed substrate (B $\rightarrow$ E, process no.~5) or the covered substrate (F $\rightarrow$ E, process no.~6).

As shown in Fig.~S1 of the Supplemental Material, a metastable site M exists beneath the graphene island edge and acts as an intermediate state connecting sites B and F. Consequently, edge attachment can proceed through both direct pathways and indirect pathways involving the metastable site M. To incorporate these coupled edge-transfer processes into an effective kinetic description, we employ a star--delta (Y--$\Delta$) transformation adapted from electrical circuit theory, as illustrated in the inset of Fig.~S1. By assuming the
coverage of metastable site M achieves steady state, the effective hopping rates are obtained as

\begin{eqnarray}
\nu_2 = \frac{ \nu_{\text{BM} } \nu_{\text{MF} } }{\nu_{\text {M-tot}} }, \quad 
\nu_5 = \frac{ \nu_{\text{BM} } \nu_{\text{ME} } }{\nu_{\text {M-tot}} } + \nu_{\text{BE} }, \nonumber \\
\nu_3 = \frac{ \nu_{\text{FM} } \nu_{\text{MB} } }{\nu_{\text {M-tot}} }, \quad
\nu_6 = \frac{ \nu_{\text{FM} } \nu_{\text{ME} } }{\nu_{\text {M-tot}} } + \nu_{\text{FE} }, \label{mu2356}
\end{eqnarray}
where $\nu_{\text {M-tot}}=\nu_{\text{MB} }+\nu_{\text{MF} }+\nu_{\text{ME} }$. The hopping rate of the elementary process from site $i$ to $j$ is calculated by $\nu_{ij}=\nu_0 \exp (-\beta \Delta G_{ij} )$ with $\beta=(k_BT)^{-1}$. The direct pathway $\nu_{\mathrm{FE}}$ is neglected because the corresponding transition is sterically blocked by the graphene edge.

To investigate strain effects, a uniform biaxial strain $\varepsilon=(a-a_0)/a_0$ is applied to the Cu(111) slab, where $a_0=3.634$~\AA\ is the optimized lattice constant of bulk Cu at 0~K. Both tensile and compressive strain up to $2\%$ are considered. The computational setup and first-principles methodology follow those established in our previous work\cite{Wangtt-PRM-2025}.

\section{Strain-Dependent Energetics and Kinetics}
\label{S3}

The growth mode of graphene is governed by the competition between first-layer lateral expansion and second-layer nucleation. To establish how substrate strain modifies this competition, we systematically investigate both the thermodynamic and kinetic aspects of graphene growth on Cu(111). From the thermodynamic perspective, strain-dependent growth driving forces and cluster formation energies are evaluated to characterize the nucleation energy landscape. From the kinetic perspective, the strain dependence of surface diffusion and attachment processes is analyzed to determine how strain modifies the transport kinetics of carbon monomers. Together, these thermodynamic and kinetic effects provide the microscopic foundation underlying the generalized growth phase diagram.

\subsection{Thermodynamic Analysis: Growth Driving Forces and Nucleation Barriers}
\label{S3-1}

The thermodynamic driving force for graphene growth is determined by the carbon supersaturation, defined as the chemical potential difference between surface carbon monomers and graphene on the same substrate, $\Delta\mu = \mu_{\text{monomer}} - \mu_{\text{gr}}$, where the monomer chemical potential is given by $\mu_{\mathrm{monomer}}=\mu_{\ce{CH4}}-2\mu_{\ce{H2}}$. Within the ideal-gas approximation, the supersaturation can be expressed as  $\Delta\mu = \Delta\mu^\ominus(T) - kT\ln \frac{ (P_{\ce{H2}})^2 }{P^{\ominus} P_{\ce{CH4}} }$, where $\Delta\mu^\ominus(T)=\mu^\ominus_{\ce{CH4}}-2\mu^\ominus_{\ce{H2}}-\mu_{\mathrm{gr}}$ represents the supersaturation under the standard pressure $P^\ominus=1$~bar. The temperature dependence of $\Delta\mu^\ominus(T)$ can be further decomposed into enthalpic and entropic contributions according to $\Delta\mu^\ominus(T)=\Delta h^\ominus-T\Delta s^\ominus$. Using the experimentally determined values $\Delta h^\ominus=0.9515$~eV and $\Delta s^\ominus=1.119\times10^{-3}$~eV/K for graphene growth on Cu\cite{Leidinger-JPCC-2021}, the supersaturation can be evaluated as a function of the experimental parameters $t=P_{\ce{H2}}/P^\ominus$ and $w=P_{\ce{H2}}/P_{\ce{CH4}}$. As summarized in Table~S1 and illustrated in Fig.~S2, $\Delta\mu$ can reach values as high as $\sim1.5$~eV at low $t\times w$\cite{Benjamin-CM-2017,LeeSeunghyun-NanoLett-2010}. Increasing $t\times w$ monotonically reduces the supersaturation and eventually drives the system into the etching regime with $\Delta\mu<0$\cite{ChoSeongYong-Nanoscale-2015}. Note that the chemical potential of carbon atom in graphene on Cu substrate is expected to become higher at enhanced temperature due to the increasing lattice mismatch between graphene and Cu substrate, which leads to a smaller driving force for the same partial pressure conditions than in graphite.

\begin{figure}[b]
\centering
\includegraphics[width=\linewidth]{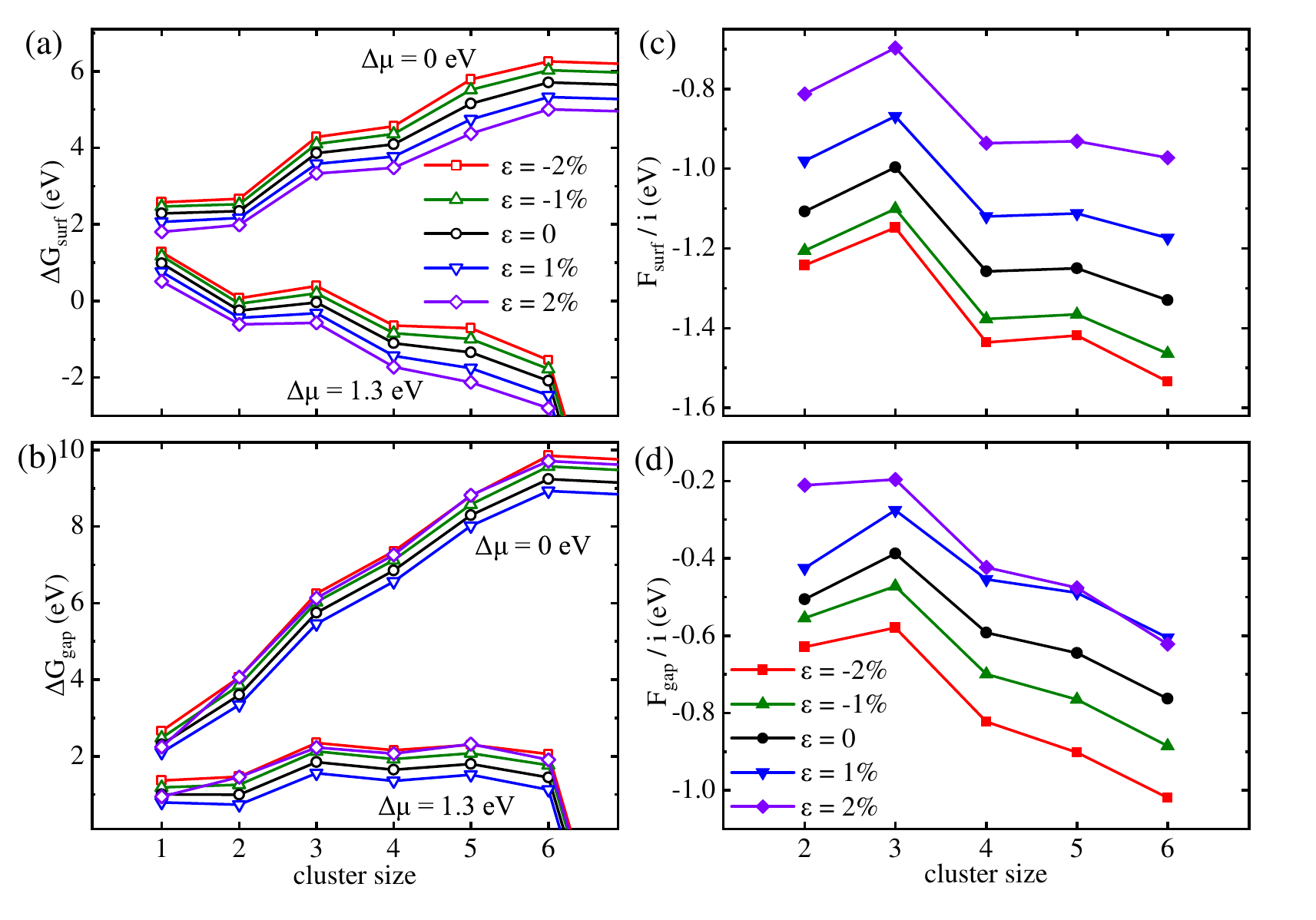}
\caption{The formation works of C clusters (a) on the exposed substrate and (b) at the graphene/substrate interface under different strains. The free energy change per C atom for C cluster formation (c) on the exposed substrate and (d) at the graphene/substrate interface under different strains.}
\label{F2}
\end{figure}

Beyond the thermodynamic driving force, we next examine the formation of carbon clusters and the associated nucleation behavior. The formation work of a carbon cluster C$_i$ containing $i$ atoms is defined as $\Delta G(i) = E_{a}(i) -k_B T \ln q_i - i(\mu_{\text{gr}} + \Delta\mu)$, where $E_a(i) = E_{\text{tot}}(i) - E_{\text{sub}}$ is the energy difference between cluster-adsorbed substrate and bare substrate, and $q_i$ is the partition function of the cluster confined within a single adsorption site. According to the critical condition $\partial\Delta G(i)/\partial i|_{i=i^*} = 0$ in classical nucleation theory, the critical size $i^*$ generally decreases with increasing $\Delta \mu$. 
In the present work, the entropic contribution is neglected for simplicity, yielding $\Delta G(i) \approx E_a(i) - i E_{\text{gr}} - i \Delta\mu$, where $E_{\text{gr}}$ is the energy per carbon atom in graphene on the same substrate.

Using the most stable linear configurations of C$_i$ clusters ($i=1\sim6$), we calculate the formation work on both the exposed substrate ($\Delta G_{\mathrm{surf}}$) and the graphene-covered substrate interface ($\Delta G_{\mathrm{gap}}$), as shown in Fig.~\ref{F2}(a,b). For all cluster sizes considered here, $\Delta G_{\mathrm{surf}}$ remains lower than $\Delta G_{\mathrm{gap}}$, indicating that first-layer nucleation is thermodynamically more favorable than second-layer nucleation. This energetic preference is further reflected in the corresponding critical nucleus sizes. For $\Delta\mu=1.3$~eV, the critical nucleus size is $i^*=1$ for first-layer nucleation but increases to $i^*=3$ for second-layer nucleation. The strain dependence of $\Delta G(i)$ further reveals distinct responses for the two growth layers. On the exposed Cu surface, $\Delta G_{\mathrm{surf}}(i)$ generally decreases under tensile strain, indicating an enhanced thermodynamic driving force for first-layer nucleation. In contrast, the strain dependence of $\Delta G_{\mathrm{gap}}(i)$ exhibits a notable anomaly under $2\%$ tensile strain. This behavior is likely associated with the reduced graphene--substrate interfacial spacing induced by strong tensile deformation. 

According to classical nucleation theory\cite{Walton-JCP-1962}, the equilibrium coverage of an $i$-atom cluster is given by $\theta_{\ce{C}_i}=\exp[-\beta \Delta G(i)]$, while the monomer coverage satisfies $\theta_{\ce C}=\exp[-\beta \Delta G(1)]$. The cluster coverage can therefore be rewritten as $\theta_{\ce{C}_i} =(\theta_{\ce{C}})^i \exp{[ -\beta F(i)]}$, where $F(i)=\Delta G(i)-i \Delta G(1)$ is the free energy cost of forming a cluster $\ce{C}_i$ from $i$ surface monomers. Since $F(i)$ directly governs the equilibrium cluster population and the steady-state nucleation rate, we further analyze the strain dependence of the formation energy per atom, $F(i)/i$ for cluster size of $i=1$--$6$. The corresponding results for first-layer clusters ($F_{\mathrm{surf}}/i$) and second-layer clusters ($F_{\mathrm{gap}}/i$) are shown in Fig.~\ref{F2}(c) and (d), respectively. For simplicity, entropic contributions are again neglected, yielding $F(i)=\Delta E_a(i)-i \Delta E(1)$.

The calculated formation energies indicate a clear energetic preference for first-layer nucleation over second-layer nucleation. Specifically, $F_{\mathrm{surf}}$ remains consistently lower than $F_{\mathrm{gap}}$, indicating that cluster formation on the exposed Cu surface is thermodynamically more favorable than beneath the graphene overlayer. Overall, both $F_{\mathrm{surf}}/i$ and $F_{\mathrm{gap}}/i$ generally decrease with increasing cluster size, reflecting the enhanced stability of larger carbon clusters. A notable exception occurs for the dimer, whose formation energy per atom is lower than that of the trimer owing to the exceptionally strong C--C covalent bonding within the dimer configuration. In addition, both $F_{\mathrm{surf}}$ and $F_{\mathrm{gap}}$ increase under tensile strain, suggesting that compressive strain favors cluster nucleation.

The opposite strain dependences of $\Delta G(i)$ and $F(i)$ indicate a size-dependent response of $E_a(i)$ to substrate strain, since $\Delta G(i)$ and $F(i)$ are referenced to $E_{\text{gr}}$ and $E_a(1)$, respectively. Indeed, for clusters with size $i=1$--$6$, $E_a(i)$ decreases monotonically with increasing tensile strain, while the magnitude of this reduction becomes weaker for larger clusters. In graphene, however, the energy per carbon atom $E_{\mathrm{gr}}$ increases under tensile strain. This behavior originates from the competition between two opposing effects: tensile strain strengthens the interaction between carbon clusters and the Cu substrate, while simultaneously inducing substantial deformation of the C--C bonding network, which destabilizes carbon aggregation.

\subsection{Kinetic Processes: Surface Diffusion and Attachment Dynamics}
\label{S3-2}

As illustrated schematically in Fig.~\ref{F1}, both the precursor decomposition stage and the surface growth stage involve multiple thermally activated elementary processes. Because methane decomposition occurs locally at individual adsorption sites, the associated reaction barriers are expected to exhibit only weak sensitivity to substrate strain. In contrast, the surface growth stage involves long-range monomer transport and edge attachment processes that are intrinsically coupled to the substrate lattice geometry. We therefore focus here on the strain dependence of the surface diffusion and attachment kinetics that govern graphene growth.

The NEB energy profiles of the elementary hopping processes are shown in Fig.~S3 for both unstrained Cu(111) and a representative tensile strain of $2\%$. The most striking result is that the minimum energy pathway for diffusion on the exposed substrate (process no.\ 1) is altered under tensile strain. As shown in Fig.\ S4, when compressive strain is applied, the optimal pathway hopping between two equilibrium fcc sites is nearly parallel to the [111] direction, denoted as the "nosym" pathway. When tensile strain is applied, however, the pathway automatically relaxes to a high-symmetry path, where the saddle point resides precisely at an hcp site, denoted as the "sym" pathway.

\begin{figure}[!b]
\centering
\includegraphics[width=\linewidth]{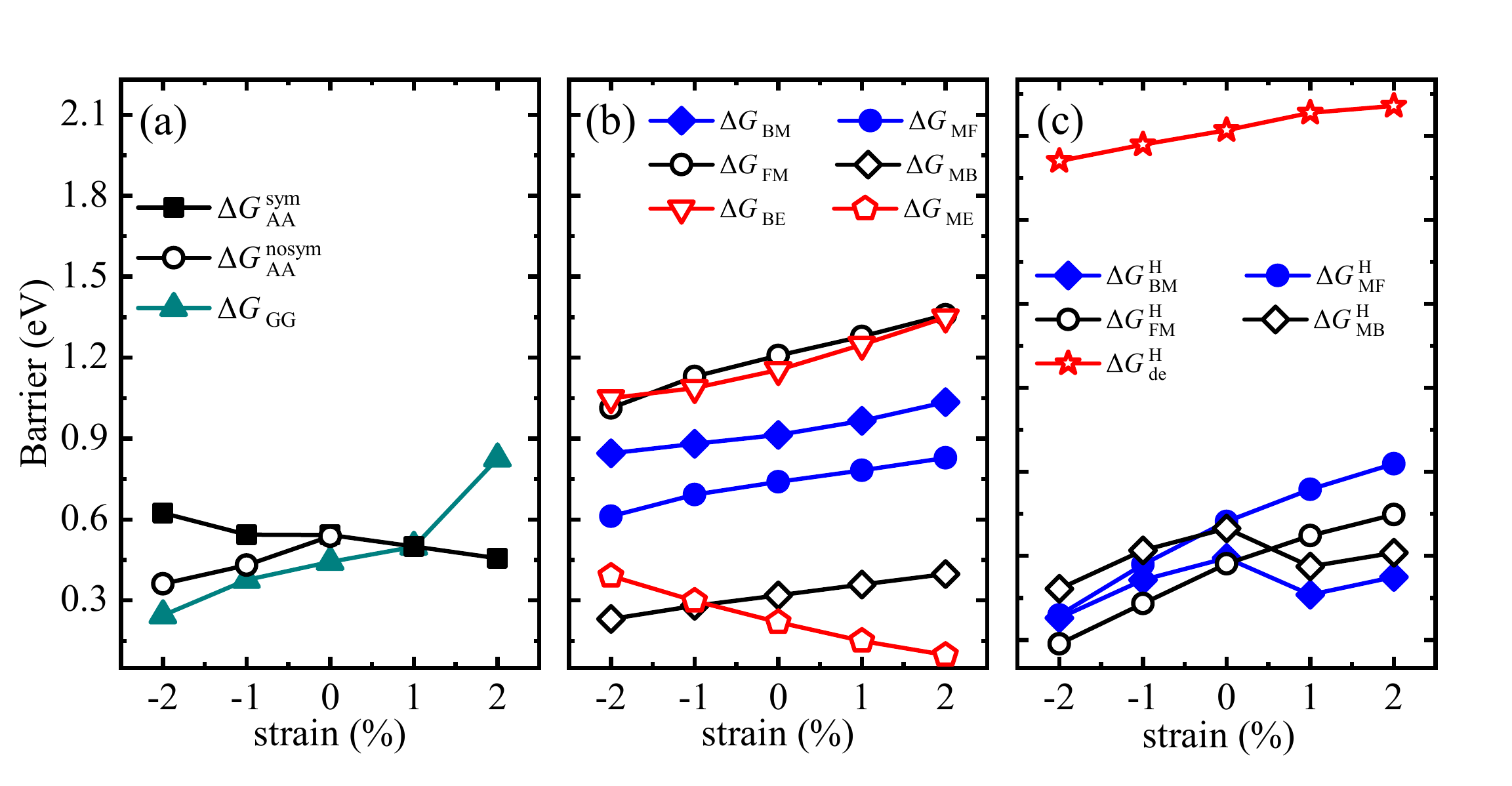}
\caption{The change in surface process barriers under different strains. (a) The surface diffusion processes far away from the graphene island edge. (b) The surface diffusion processes across the island edge and the attachment process with the freestanding edge. (c) The surface diffusion processes across the island edge and the H detachment process with the H-terminated edge.}
\label{F3}
\end{figure}

The activation barriers for all the elementary kinetic processes were systematically calculated under four strain conditions ($-2\%$, $-1\%$, $1\%$, and $2\%$), with the results summarized in Fig.\ \ref{F3}(a) and (b). The two diffusion pathways associated with process no.~1 exhibit opposite strain dependences. Compressive strain reduces $\Delta G_{\text {AA} }^{\text{nosym} }$ but enhances $\Delta G_{\text {AA} }^{\text{sym} }$, thereby rendering the "nosym" pathway kinetically preferred. In contrast, the "sym" pathway becomes dominant when tensile strain is applied. Consequently, the diffusion barrier on the exposed substrate, $\Delta G_{\mathrm{AA}}$, decreases under both compressive and tensile strain, irrespective of the strain sign. For diffusion across the island edge or beneath the island, the "nosym" pathway remains energetically favored across the entire strain range considered here. Therefore, the activation barriers of these processes (B $\leftrightarrow$ M $\leftrightarrow$ F, G $\leftrightarrow$ G) are reduced by compressive strain and raised by tensile strain. The direct attachment to the edge (B $\rightarrow$ E) also follows a similar trend, but the attachment barrier from the metastable site M displays an opposite trend that decreases with increasing strain.

For comparison, we also consider the influence of hydrogen termination at the island edges on these kinetics, which is experimentally relevant under high $\ce{H2}$ partial pressures. As shown in Fig.~\ref{F3}(c), hydrogen-terminated edges evidently reduce the barriers for edge crossing, namely $\Delta G_{\mathrm{BM}}$, $\Delta G_{\mathrm{MF}}$, and their reverse processes. Similar to the H-free case, these barriers generally increase under tensile strain and decrease under compressive strain. A notable difference is the emergence of a non-monotonic strain dependence when the system evolves from compressive to tensile strain, which most likely originates from strain-induced shifts of the diffusion pathway, analogous to the behavior observed for $\Delta G_{\mathrm{AA}}$. Meanwhile, hydrogen termination strongly suppresses edge attachment because the hydrogen detachment barrier $\Delta G_{\mathrm{de}}^{\mathrm H}$ ($\sim 2$ eV) is substantially higher than both $\Delta G_{\mathrm{BE}}$ and $\Delta G_{\mathrm{ME}}$. 

\section{Enhanced Rate-Equation Framework}

\subsection{Gas-phase reactions: from CVD to quasi-PVD}
\label{S4-1}
To establish a tractable rate-equation description of graphene CVD growth, the complex multistep precursor reaction is transformed into effective deposition and desorption processes. Within this treatment, the overall CVD growth can be mapped onto an effective quasi-PVD framework.
The decomposition of methane ($\mathrm{CH}_4$) precursors occurs through either direct gas-phase dissociation or catalytic dehydrogenation on the Cu surface following adsorption. In both situations, the decomposition process is modeled as a sequence of four elementary dehydrogenation steps. Specifically, the elementary steps and the overall reactions of CH$_4$ decomposition on Cu are as follows,
\begin{align}
\ce{CH_{4-l}(a) + {\mathbb A} &\rightleftharpoons CH_{3-l}(a) + H(a)},~~ (l=0,1,2,3) \tag{R1} \\
\ce{CH_4(a) + 4 {\mathbb A} &\rightleftharpoons C(a) + 4H(a)}. \tag{R2}
\end{align}
Here ${\mathbb A}$ denotes an empty adsorption site of the Cu substrate and (a) indicates the adsorbed state. The kinetics of the $(l+1)$-th dehydrogenation step (R1) are characterized by the forward and backward rate constants, denoted as $k_{l}^+$ and $k_{l+1}^-$, respectively. Meanwhile, the adsorption and desorption rate constants of CH$_{4-l}$ ($l=0,1,2,3$) are denoted as $f_{l}$ and $z_{l}$. Accordingly, the surface density of the intermediate $\ce{CH}_{4-l}$, denoted as $\theta_l$, follows the evolution equation $\partial_t \theta_l = ( P_l f_l \theta_{\mathbb A} - z_l \theta_l) + (\theta_{l-1} k_{l-1}^{+} \theta_{\mathbb A} - \theta_{l} k_{l}^{-} \theta_{\ce{H}} ) - (\theta_{l} k_{l}^{+} \theta_{\mathbb A} - \theta_{l+1} k_{l+1}^{-} \theta_{\ce{H}} )$. Here $P_l$ represents the partial pressure of the gas-phase species, while $\theta_{\mathbb A}$ and $\theta_{\mathrm H}$ denote the densities of empty adsorption sites and adsorbed hydrogen atoms on the substrate, respectively. Under typical CVD conditions, the first term (direct adsorption and desorption) of the evolution equation is negligible compared with the other two terms (dehydrogenation flux). Therefore, assuming the dehydrogenation steps reach steady state yields an expression for the effective decomposition rate $k_d^+$ of CH$_4$ in the overall reaction (R2),
\begin{equation}
\frac{1}{\theta_{\mathbb A}^3 k_d^+} = \frac{1}{ k_0^+} + \sum_{i=1}^{3} \frac{1}{K_{0i} k_i^+}, \quad
K_{0i} = \prod_{l=0}^{i-1} \frac { k_l^{+} \theta_{\mathbb A}} {k_{l+1}^{-} \theta_{\ce{H}} }.
\end{equation}
Similarly, the effective backward rate constant of (R2) is given by $k_d^{-} = K_{04} k_d^{+} \theta_{\mathbb A}^4 / \theta_{\ce{H}}^4 $, where $K_{04}$ is defined analogously to the $K_{0i}$ above. The resulting effective rate constant $k_d^+$ and $k_d^-$ therefore represents the net methane decomposition flux after coarse-graining over the intermediate hydrocarbon species.

Further assuming that the surface coverage of $\ce{CH4}$ reaches a steady state via the overall decomposition reaction (R2) much faster than the growth timescale of carbon islands, the multistep CVD chemistry can be reduced to an effective rate equation governing the surface concentration of carbon monomers, coverage $\theta_{\ce{C}}$,
\begin{equation}
\frac{d\theta_{\ce{C}}}{dt} = F_{\text{eff}} \theta_{\mathbb A} - (k_g + z_{\text{eff}}) \theta_{\ce{C}}, \label{eq:Crate}
\end{equation}
where $k_g$ denotes the effective consumption rate constant of carbon monomers due to nucleation and island growth, as will be specified later.
The effective deposition flux $F_{\text{eff}}$ and the effective desorption rate constant $z_{\text{eff}}$ of carbon monomers are given as follows,
\begin{align}
F_{\text{eff}} &= P_{\ce{C}} f_{\ce{C}} + \frac{P_{\ce{CH4}} f_{\ce{CH4}} k_d^{+} \theta_{\mathbb A}^4 }{z_{\ce{CH4}} + k_d^{+} \theta_{\mathbb A}^4 }, \label{eq:Fs}\\
z_{\text{eff}} &= z_{\ce{C}} + \frac{z_{\ce{CH4}} k_d^{-} \theta_{\ce{H}}^4 }{z_{\ce{CH4}} + k_d^{+} \theta_{\mathbb A}^4 }. \label{eq:zs}
\end{align}
The first term in $F_{\text{eff}}$ accounts for direct impingement of C atoms (negligible under typical CVD conditions), and the second term represents the net supply from CH$_4$ decomposition. Similarly, the desorption rate $z_{\text{eff}}$ contains a physical desorption term $z_{\ce{C}}$ (usually very small due to the strong chemisorption of carbon on Cu) and a chemical desorption term arising from the reverse dehydrogenation in (R2).  $z_{\text{eff}}$ is dominated by the reverse dehydrogenation term at high $P_{\ce{H2}}$. This chemical desorption pathway, analogous to the mechanism quantified by Vlassiouk \emph{et al.}\cite{Vlassiouk-ACSNano-2011}, effectively depletes carbon monomers and reduces their steady-state concentration. Therefore, $z_{\text{eff}}$ represents an effective non-growth depletion for carbon monomers. In this quasi-PVD framework, Eq. (\ref{eq:Crate}) is analogous to a PVD rate equation but with renormalized deposition and desorption rates that incorporate chemical reactions. 

\subsection{First-layer nucleation and growth}
\label{S4-2}
Based on classical nucleation theory\cite{Venables1973, Venables1984}, first-layer clusters are treated as stable islands only when their size $n$ exceeds a critical value $i^*$, otherwise, they remain as unstable embryos.  Assuming a steady-state formation current for embryos of sizes up to $n\leq i^*$, we obtain the steady-state nucleation rate
\begin{equation}
J_{\text{nuc}}(i^*) = w_{i^*}^{+} \theta_{\ce{C}}^{i^*+1} \exp{[-\beta F_{\text{surf}}(i^*) ]}, \label{eq:nuc}
\end{equation}
where $w_{i^*}^{+}=\sigma_{i^*}\nu_1 $ is the forward reaction rate constant for the attachment process $\ce{C + C_{i^*} \rightleftharpoons C_{i^*+1} }$, with $\sigma_{i^*}$ the capture number of critical nucleus, and $\nu_1=\nu_\text{AA}$ the rate of process no.~1 (diffusion of carbon monomers on the exposed substrate).

By grouping all stable islands of size $n>i^*$ together, with a total density of first-layer stable islands $\theta_{\text{isl}}$, an average capture number $\sigma_{s}$ and an average attachment rate to the island edge $w_s^+ = \sigma_{s}\nu_1$, the total monomer flux toward these stable islands is given by $J_{\text{isl}} =w_s^{+} \theta_{\ce{C}} \theta_{\text{isl}}$. The total consumption rate of C monomers, accounting for both nucleation and island growth, is thus defined as $k_g \theta_{\ce{C}} =(i^*+1) J_{\text{nuc}}+ J_{\text{isl}}$. In the growth regime, $k_g \theta_{\ce{C}} \simeq J_{\text{isl}}$ since $(i^*+1)J_{\text{nuc}} \ll J_{\text{isl}}$.

If desorption is negligible ($z_{\text{eff}}\to0$), the monomer population is governed solely by the competition between deposition and island capture, $\theta_{\ce{C}} = F_{\text{eff}}\theta_{\mathbb A}/( \sigma_s \nu_1\theta_{\text{isl}})$, according to Eq.\ (\ref{eq:Crate}). This leads to the familiar PVD scaling for island density. In contrast to conventional PVD growth, graphene CVD involves an additional non-growth depletion pathway arising from chemical desorption. Consequently, the effective desorption rate $z_{\text{eff}}$ must be retained explicitly in Eq.~(\ref{eq:Crate}). Accounting for this depletion pathway, the steady-state monomer density is expressed as
\begin{equation}
\theta_{\ce{C}} = F_{\text{eff}} \theta_{\mathbb A}/ (\sigma_s \nu_1 \theta_{\text{isl}} + z_{\text{eff}}). \label{eq:mono}
\end{equation}
Substituting Eq.\ (\ref{eq:mono}) into (\ref{eq:nuc}), and then integrating the nucleation rate $J_{\text{nuc}}$, we obtain the saturation island density\cite{Wangtt-PRM-2025},
\begin{equation}
\theta_{\text{isl}} = \left[ B_{i^*} \Gamma^{-i^*} + Z^{i^* +2} \right]^{1/(i^* +2)} - Z, \label{eq:theta_s}
\end{equation}
where $B_{i^*} = \frac{(i^* +2) \sigma_i}{\sigma_s^{i^* +1}} \Theta_c \exp{[-\beta F_{\text{surf}}({i^*})]}$, and
\begin{equation}
\Gamma = \nu_1 / F_{\text{eff}},\quad Z = z_{\text{eff}}/( \sigma_s \nu_1). \label{eq:Gamma}
\end{equation}
Here $\Theta_c$ is the coverage at which island saturation occurs. Within this generalized framework, $\Gamma$ characterizes the competition between surface diffusion and deposition, while $Z$ quantifies the relative strength of chemical desorption against island growth. 
Notice that when $Z=0$, Eq.\ (\ref{eq:theta_s}) reduces to the familiar PVD scaling $\theta_{\text{isl}} \propto \Gamma^{-i^* /(i^* +2)}$.

\subsection{Second-layer nucleation beneath islands}
\label{S4-3}
As first-layer graphene islands expand, carbon monomers on the exposed Cu substrate can diffuse across the island edges and accumulate beneath the graphene islands. The monomers beneath the first layer provide the precursor reservoir for second-layer nucleation. To distinguish the monomer populations outside and beneath the islands, we denote the corresponding monomer concentrations by $\eta_1$ and $\eta_2$, respectively. With the effective deposition rate $F_{\text{eff}}$, the condition of mass conservation gives
\begin{equation}
\pi (L^2 - R^2) F_{\text{eff}} \frac{\theta_{\text{isl}}}{\theta_{\text{isl}} + Z}=2\pi R(\eta_1 \nu_5 + \eta_2 \nu_6), \label{eta12}
\end{equation}
where $L = 1/\sqrt{\pi \theta_{\text{isl}}}$ is half the average island separation. The factor $\theta_{\text{isl}}/(\theta_{\text{isl}} + Z)$ accounts for the competition between island growth and monomer depletion through chemical desorption. Assuming a steady state for $\eta_2$ yields the kinetic relation $\eta_1 \nu_2 =\eta_2 (\nu_3+\nu_6) $. Substituting this relation into Eq.\ (\ref{eta12}), we derive the monomer density beneath the island
\begin{equation}
\eta_2 = \frac{L^2 - R^2}{2R \alpha \Gamma} \frac{\theta_{\text{isl}}}{\theta_{\text{isl}} + Z}, \label{eq:theta1}
\end{equation}
where $\alpha$ accounts for the competition between the kinetic processes,
\begin{equation}
\alpha = (\nu_3\nu_5 + \nu_2\nu_6 + \nu_5\nu_6)/(\nu_1\nu_2). \label{eq:alpha}
\end{equation}
Here $\nu_{2,3,5,6}$ denote the effective kinetic rates obtained from the Y--$\Delta$ transformation in Eq.~(\ref{mu2356}), while $\nu_1=\nu_{\text{AA}}$ and $\nu_4=\nu_{\text{GG}}$ correspond to monomer diffusion on the exposed and island-covered substrate, respectively. 

Because no direct deposition occurs beneath the first-layer graphene islands, the confined monomer population rapidly reaches a spatially uniform concentration. The corresponding nucleation rate for a second-layer nucleus with critical size $i^*$ is therefore $J_{\text{gap}} (i^*)= \sigma_{i^*} \nu_4 \eta_2^{i^*+1} \exp{[-\beta F_{\text{gap}}(i^*)]}$, and the total number of second-layer nuclei that form before the first-layer islands coalesce ($R$ reaches $L$) is
\begin{equation}
I_{\text{gap}}(i^*) = \int_{R_n}^{L} J_{\text{gap}}(i^*) \pi R^2 \frac{dt}{dR} dR,
\end{equation}
where $R_n$ represents the minimum radius at which second-layer nucleation can occur. This geometric threshold is associated with a 7-ring $\ce{C24}$ cluster, corresponding to $R_n = 3$ in units of $\sqrt{A_0}$, where $A_0$ represents the area of a graphene unit cell. Substituting Eqs.~(\ref{eq:theta1}) into the nucleation rate $J_{\text{gap}}$, and recognizing Eq.~(\ref{eta12}) as the growth rate $ \partial_t (\pi R^2)$, we obtain
\begin{equation}
I_{\text{gap}}(i^*) = \frac{C_{i^*} g_{i^*}(L)}{\alpha (2\alpha \Gamma)^{i^*}} \left( \frac{\theta_{\text{isl}}}{\theta_{\text{isl}} + Z} \right)^{i^*}, 
\label{eq:I_i}
\end{equation}
where $C_{i^*} = \pi \lambda \sigma_{i^*} \exp{[-\beta F_{\text{gap}}(i^*)]}$, $\lambda = \nu_4/\nu_1$, and $g_{i^*}(L) = \int_{R_n}^{L} (L^2 - R^2)^{i^*} R^{2-i^*} dR$.

The growth mode is determined by comparing $I_{\text{gap}}(i^*)$ with unity. 
If $I_{\text{gap}}(i^*) < 1$, no second-layer nucleus forms before island coalescence, resulting in SLG growth. Conversely, if $I_{\text{gap}}(i^*) > 1$, second-layer nuclei can form prior to coalescence, thereby enabling BLG growth. The condition $I_{\text{gap}}(i^*) = 1$ therefore defines the phase boundary in the $(\alpha,\Gamma)$ parameter space, with chemical desorption incorporated through the additional kinetic parameter $Z$.

\section{Effects of Strain and Chemical Desorption on Graphene Growth}

To establish a more realistic description of CVD growth, the phase diagram is generalized by incorporating the effects of substrate strain and chemical desorption. As discussed in Sec.~\ref{S3}, substrate strain strongly influences both cluster formation energetics and surface kinetics. During growth, such strain naturally arises from either lattice mismatch in epitaxial systems\cite{TangJilin-NanoResearch-2023} or thermal expansion at elevated temperatures. In the present work, we focus on the latter mechanism and use temperature as the experimentally accessible variable that indirectly controls strain through thermal expansion of the Cu substrate. The temperature-dependent tensile strain is determined from the thermal expansion of Cu relative to its lattice constant at 0~K. The diffusion barriers and cluster formation energies at finite temperature are then obtained by extrapolating the corresponding DFT results to the strain state associated with that temperature (see Figs.~S5 and S6 for details). 
	
Thermal strain affects the growth phase behavior through two concurrent mechanisms: it both modifies the kinetic parameters $\alpha$ and $\Gamma$ and shifts the phase boundary through strain-dependent cluster formation energies. In addition, chemical desorption introduces a finite parameter $Z$, providing an additional kinetic pathway for regulating graphene growth. In the following, we first analyze the temperature dependence of $\alpha$, $\Gamma$, and $Z$, and then construct the generalized phase diagram. Finally, based on this framework, we discuss possible strategies for controlled BLG growth.

\subsection{Variation of Parameters $\alpha$, $\Gamma$ and $Z$}
\label{S5-1}
To evaluate the kinetic parameters $\Gamma$ and $Z$ defined in Eq.~\eqref{eq:Gamma}, the surface coverages of vacancies ($\theta_{\mathbb A}$) and atomic hydrogen ($\theta_{\ce H}$) must first be determined. This is achieved by incorporating the adsorption and dissociation kinetics of hydrogen molecules together with reactions (R1) and (R2),
\begin{align}
	\ce{H2(g) }+ {\mathbb A} \rightleftharpoons \ce{H2(a)}, \quad \ce{ H2(a)} + {\mathbb A} \rightleftharpoons 2 \ce{ H(a) }. \tag{R3}
\end{align}
Here, $f_{\ce{H2}}$ and $z_{\ce{H2}}$ denote the adsorption and desorption rate constants of $\ce{H2}$, while $k_{\ce H}^{+}$ and $k_{\ce H}^{-}$ represent the forward and backward rate constants for hydrogen dissociation on the substrate. 
The coverages of molecular hydrogen ($\theta_{\ce{H2}}$) and atomic hydrogen ($\theta_{\ce H}$) are obtained from the steady-state conditions of reactions (R1)--(R3), while $\theta_{\mathbb A}$ satisfies the conservation constraint. In a typical CVD growth, the partial pressure of $\ce{H2}$ is much larger than that of $\ce{CH4}$, rendering the occupancy of hydrocarbon species ($\ce{CH}_i$) negligible compared to hydrogen. Consequently, the site-conservation constraint simplifies to $\theta_{\ce{H2}} + \theta_{\ce{H}} + \theta_{\mathbb{A}} \approx 1$. The hydrogen coverage $\theta_{\ce{H}}$ is predominantly governed by the dissociative adsorption of $\ce{H2}$. Under these steady-state conditions, the coverages are derived as $\theta_{\mathbb A}=\frac{1}{ 1+ P_{\ce{H2}} f_{\ce{H2}} / z_{\ce{H2}} + \sqrt{ P_{\ce{H2}} f_{\ce{H2}} k^{+}_{\ce{H}} / (z_{\ce{H2}} k^{-}_{\ce{H}} ) } }$ and $\theta_\mathrm{H} =\theta_{\mathbb A} \sqrt{ P_{\ce{H2}} f_{\ce{H2}} k^{+}_{\ce{H}} / (z_{\ce{H2}} k^{-}_{\ce{H}} ) }$.

According to the kinetic theory of gases, the deposition rate constant is given by $f = a^2 (2\pi m k_B T)^{-1/2}$, where $a^2$ denotes the area of an adsorption site on the Cu(111) surface and $m$ is the molecular mass. The partial pressure of gas-phase carbon monomers is assumed to be negligible, such that the first term in Eq.~\eqref{eq:Fs} can be omitted. Following transition state theory (TST), the rate constants for gas desorption ($z$) and surface reactions ($k^{\pm}$) are calculated by $(k_B T/h)\exp(-\beta \Delta G)$. The corresponding activation barriers (summarized in Table~S2) are assumed to be strain-independent because of their localized nature.

\begin{figure}[!b]
\centering
\includegraphics[width=\linewidth]{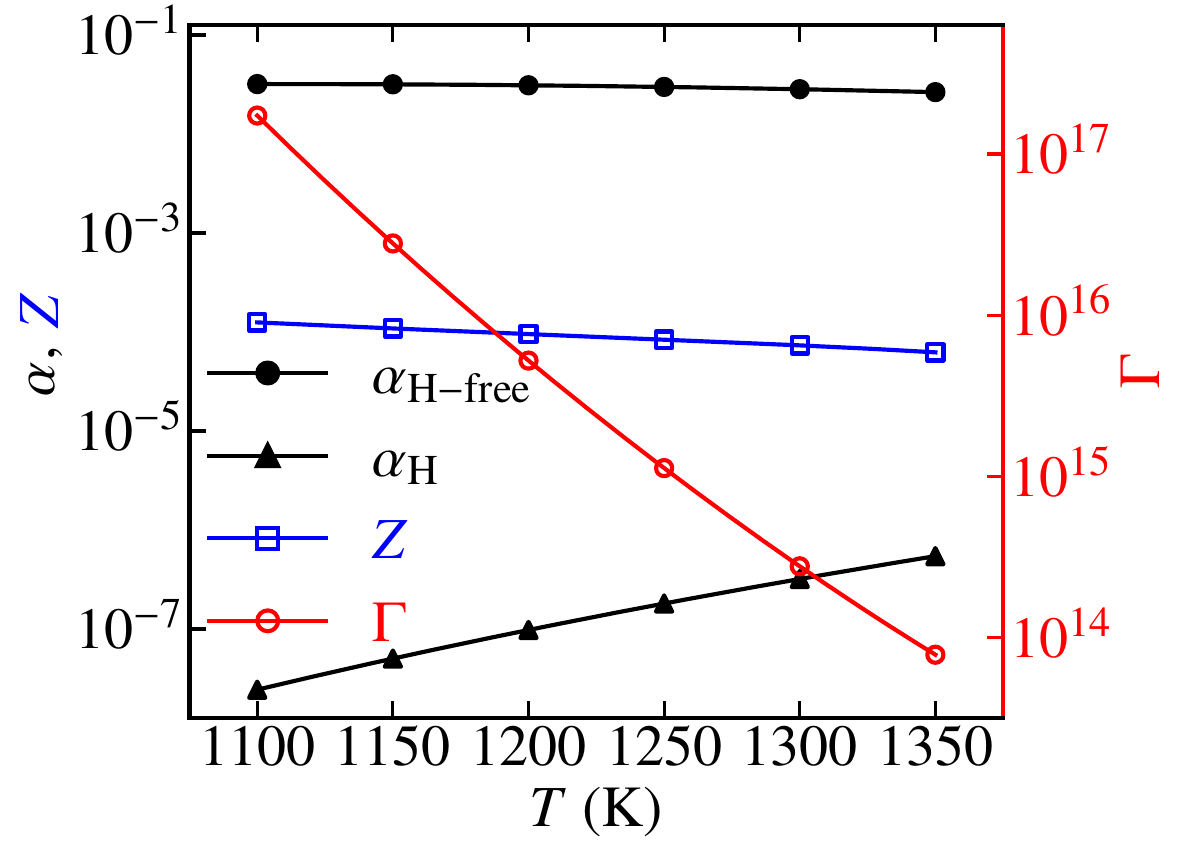}
\caption{The estimated values of parameters $\alpha$, $\Gamma$ and $Z$ as a function of temperature for $p_{\mathrm{H_2}}=0.01$ bar and $w=100$.}
\label{F4}
\end{figure}

Choosing a typical value of $P_{\ce{H2}} = 0.01\,\mathrm{bar}$ and a gas ratio $w = 100$, the temperature dependences of $\alpha$, $\Gamma$, and $Z$ are illustrated in Fig.~\ref{F4}. The values of $\alpha$ are determined for both
H-free and H-terminated graphene edges, representing the low and high $P_{\ce{H2}}$ limits, respectively. The magnitude of $\alpha$ spans a wide range from approximately $10^{-7}$ to $10^{-1}$, depending on whether the graphene island edges are hydrogen-passivated. While $\alpha_{\mathrm{H\text{-}free}}$ remains relatively stable with a slight downward trend, $\alpha_{\mathrm H}$ increases markedly with temperature. For the H-free system, the strain-induced enhancement of the effective barrier slightly outweighs the thermal activation effect as temperature increases, resulting in a weak decrease in $\alpha_{\mathrm{H\text{-}free}}$. In contrast, hydrogen termination substantially modifies the surface kinetics, with enhanced $\nu_2^{\ce H}$ and suppressed $\nu_5^{\ce H}$ and $\nu_6^{\ce H}$ relative to the H-free case. These shifts manifest as a substantially higher effective barrier, inherently resulting in a lower $\alpha_{\mathrm H}$ value. In this high-barrier regime, the exponential term becomes exceedingly sensitive to temperature fluctuations. The
gain in thermal energy ($k_BT$) thus dominates the kinetic evolution, leading to the drastic, monotonic increase in $\alpha_{\mathrm H}$ observed at elevated temperatures.

In contrast to $\alpha$, $\Gamma$ decreases monotonically with temperature and spans a wide range from $10^{14}$ to $10^{17}$ across the investigated CVD temperature window. This decrease in $\Gamma$ is dominated by the rapid increase in $F_{\text{eff}}$ at higher temperatures. The dependence of $\Gamma$ on gas partial pressures at different temperatures is further given in Fig.~S7, where a wider range of $t = P_{\ce{H2}}/P^{\ominus}$ and $w=P_{\ce{H2}}/P_{\ce{CH4}}$ is considered. Across all investigated conditions, $\Gamma$ decreases with temperature but increases with either larger $t$ or $w$, corresponding to higher $P_{\ce{H2}}$ or lower $P_{\ce{CH4}}$.

The dependence of $Z$ on growth conditions is evaluated via Eqs.~\eqref{eq:zs} and \eqref{eq:Gamma}. At fixed $P_{\ce{H2}}$, $Z$ decreases slightly with increasing temperature, as shown by the blue curve in Fig.\ \ref{F4}. Although both $\nu_1$ and $z_{\text{eff}}$ increase with temperature,  the stronger enhancement of $\nu_1$ relative to $z_{\mathrm{eff}}$ leads to an overall reduction in $Z$. In contrast, increasing in hydrogen partial pressure promotes chemical desorption, thereby yields higher values for both $z_{\text{eff}}$ and $Z$, as shown in Fig.~S8. Consequently, while thermal strain influences graphene growth through concurrent modifications of kinetic parameters, chemical desorption primarily affects the growth behavior through the evolution of the kinetic parameter $Z$. Together, these effects establish the kinetic and thermodynamic basis for the generalized phase diagram discussed below.

\subsection{Generalized Phase diagram}
\label{S5-2}

\begin{figure}[htb]
\centering
\includegraphics[width=0.8\linewidth]{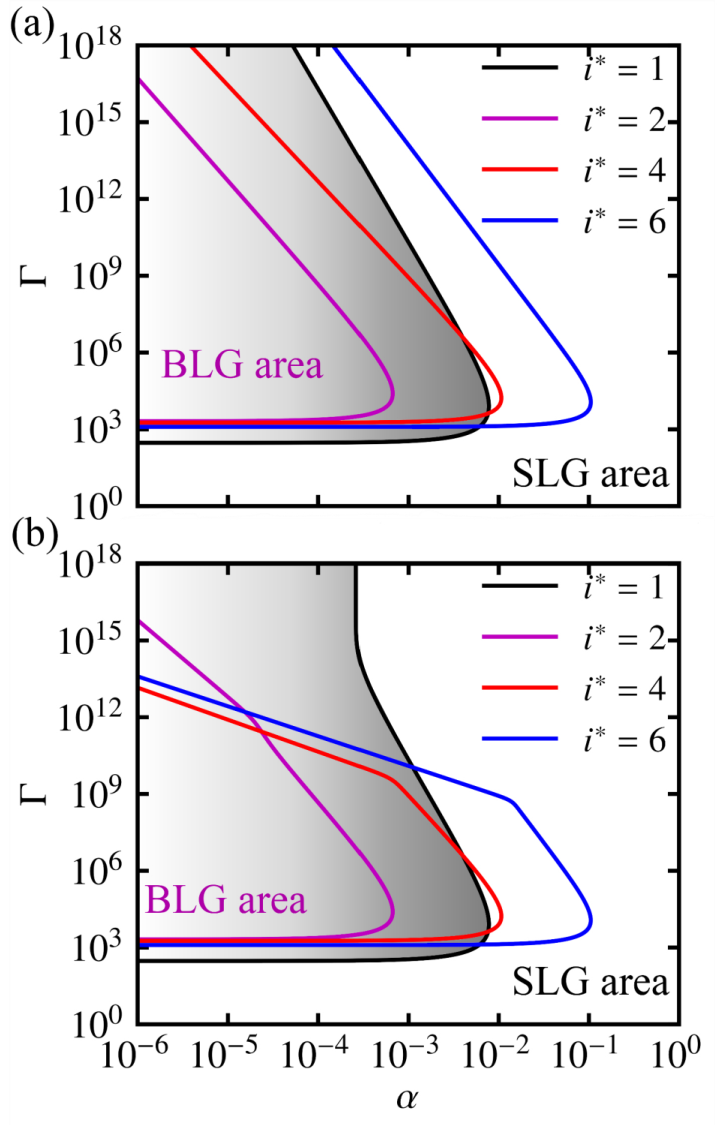}
\caption{Growth mode diagrams of graphene on Cu(111) at $T$ = 1300 K, incorporating thermal activation and substrate thermal expansion. Non-growth carbon depletion is (a) excluded ($Z=0$) and (b) included ($Z\neq 0$).}
\label{F5}
\end{figure}

We now investigate how thermal strain and chemical desorption modify the graphene growth phase behavior.
By adopting $\sigma_{i^*} = 2$, $\sigma_{s} = 10$, and $\Theta_c = 0.2$ as in our previous report\cite{Wangtt-PRM-2025}, the phase diagram is constructed via Eq.~\eqref{eq:I_i} as a function of the dimensionless parameters $\alpha$ and $\Gamma$. Figures~\ref{F5}(a) and (b) show the phase diagrams at $T = 1300$~K for $Z = 0$ and $Z \neq 0$, respectively, both incorporating the effects of thermal-expansion-induced strain.

The generalized phase diagram reveals a qualitatively different strain response between the $i^*=1$ and $i^*>1$ growth regimes. As shown in Fig.~\ref{F5}(a), the phase boundary between BLG and SLG growth deviates significantly from our previous results once temperature-dependent effects are incorporated. For $i^*=1$, the BLG growth regime contracts slightly. Since the cluster formation energy vanishes in this limit, the temperature dependence of Eq.~\eqref{eq:I_i} arises solely through the parameter $\lambda$. Thermal strain increases $\Delta G_4$ while decreasing $\Delta G_1$, thereby reducing $\lambda$ and shrinking the BLG region. In contrast, for $i^*>1$, the strain-dependent formation energies $F_{\mathrm{gap}}$ and $F_{\mathrm{surf}}$ become active simultaneously. Although the increase in $F_{\mathrm{gap}}$ suppresses second-layer nucleation, the concurrent increase in $F_{\mathrm{surf}}$ lowers the island density $\theta_{\mathrm{isl}}$ and increases the average island separation $L$, which favors second-layer growth. The competition between these two effects ultimately expands the BLG regime, particularly for larger critical nucleus sizes (e.g., $i^*=6$), representing a notable deviation from the strain-free case.

When chemical desorption is incorporated, the phase diagram is presented in Fig.~\ref{F5}(b) for $T=1300$~K and $P_{\ce{H2}}=0.01$~bar. In addition to the coefficients $B_{i^*}$ and $C_{i^*}$, the phase boundary now depends explicitly on the factor $[\theta_{\mathrm{isl}}/(\theta_{\mathrm{isl}}+Z)]^i$ in Eq.~\eqref{eq:I_i}, which couples the second-layer nucleation number to both the island coverage $\theta_{\mathrm{isl}}$ and the chemical desorption strength $Z$. In the low-$\Gamma$ regime corresponding to high deposition rates, $\theta_{\mathrm{isl}}$ remains sufficiently large such that the $Z$-dependent factor approaches unity. The resulting phase boundary therefore remains close to the $Z=0$ limit. In contrast, for large $\Gamma$, the reduced island coverage enhances the influence of chemical desorption, and the $Z$-dependent factor suppresses the second-layer nucleation number $I_i$, thereby inhibiting BLG formation. Consequently, the phase boundary bends downward in the high-$\Gamma$ regime, leading to a reduced BLG region. A notable exception occurs for $i^* = 1$, where the term $[\theta_{\text{isl}}/(\theta_{\text{isl}}+Z)]$ is insufficient to offset the favorable changes in $\ce{C}_{i^*}$ for BLG growth. As a result, the BLG domain becomes slightly expanded for large-$\Gamma$ in Fig.~\ref{F5}(b).

The temperature dependence of the phase boundaries is further examined in Figs.~S9 and S10 for the cases of $Z=0$ and $Z\neq0$, respectively. In the absence of chemical desorption ($Z=0$), the BLG region contracts slightly with increasing temperature for $i^*=1$. Since $\Delta F=0$ in this limit, the temperature dependence arises primarily from the variation of $\lambda$. In contrast, for $i^*>1$, the BLG growth window expands at elevated temperatures. This behavior originates from the competing strain-induced variations of $\Delta F_{\mathrm{surf}}$ and $\Delta F_{\mathrm{gap}}$. Although increasing temperature generally suppresses nucleation on both layers, the second-layer nucleation barrier exhibits a weaker temperature dependence than that of the first layer. Consequently, the relative suppression of first-layer nucleation becomes more pronounced, ultimately favoring BLG growth and expanding the BLG regime in phase space. For the $Z\neq0$ case, chemical desorption primarily modifies the phase boundary in the high-$\Gamma$ regime through the $Z$-dependent factor discussed above without altering the fundamental temperaturedependent evolution of the growth modes. The resulting trends therefore remain consistent with the $Z=0$ case: the BLG regime contracts slightly for $i^*=1$ but expands substantially for $i^*>1$ as temperature increases.

\subsection{Strategies for Bilayer Synthesis}

Although the phase diagram is formulated in terms of the dimensionless parameters $\alpha$ and $\Gamma$, these quantities map directly onto experimentally accessible parameters, including gas partial pressures and temperature. As summarized in Fig.~S8, $\Gamma$ typically spans the range from $10^{12}$ to $10^{20}$ under standard CVD conditions, indicating that most experiments operate within the large-$\Gamma$ regime of the generalized phase diagram. Experimentally, the hydrogen partial pressure provides an effective approach of tuning the parameter $\alpha$ through hydrogen-dependent edge kinetics. Low $P_{\ce{H2}}$ stabilizes H-free graphene edges and therefore maintains relatively large $\alpha$, favoring SLG expansion. In contrast, increasing $P_{\ce{H2}}$ promotes hydrogen termination of the graphene edge, reduces $\alpha$, and consequently facilitates BLG nucleation. More generally, these results demonstrate that layer selectivity can be achieved by mapping experimental conditions—temperature and chemical environment—onto the $(\alpha, \Gamma)$ parameter space.

More importantly, the generalized phase diagrams suggest that elevated growth temperatures can move the system deeper into the BLG-favorable region of phase space for $i^*>1$. This theoretical framework therefore provides a natural mechanistic interpretation for several experimental observations. For instance, Celebi \emph{et al.} reported that the secondary graphene flake area increases with growth temperature\cite{Celebi-NanoLett-2013}. Fabiane \emph{et al.} further observed that BLG synthesized at relatively low temperature ($\sim780\,^\circ$C) exhibits substantial structural disorder, whereas growth at higher temperatures ($\sim956$--$973\,^\circ$C) predominantly yields high-quality AB-stacked bilayer islands\cite{Fabiane-JRS-2017}. More broadly, experiments on Cu-catalyzed CVD systems consistently indicate that low-temperature growth strongly favors single-layer graphene formation\cite{Weatherup-ACSNano-2012,Mehdipour-ACSNano-2012}. These experimental trends are fully consistent with the temperature-dependent expansion of the BLG regime predicted by our generalized phase diagrams. Although low-temperature BLG synthesis remains challenging, approaches such as nucleation-density engineering or the Cu-pocket method may provide viable pathways for overcoming these kinetic limitations.

Beyond temperature-driven strain, epitaxial strain introduced during substrate preparation provides an additional route for tuning graphene growth behavior. Epitaxial tensile strain exerts a similar influence as thermal expansion, expanding the BLG growth window for $i^*>1$. This result provides a natural explanation for the enhanced bilayer yield frequently observed on heteroepitaxial substrates. For instance, the lattice mismatch between sapphire and ($\sqrt{3}\times\sqrt{3}$)Cu(111) induces an intrinsic tensile strain of approximately $6\%$ in the Cu film\cite{TangJilin-NanoResearch-2023,LiJunzhu-NatMater-2022,ZhangXuefu-small-2019,DengBing-ACSNano-2017}. As discussed in Sec.~\ref{S5-2}, such tensile strain, whether originating from epitaxial mismatch or further enhanced by thermal expansion, shifts the operating point deeper into the BLG regime of the generalized phase diagram. This cooperative effect is particularly evident in CuNi(111)/sapphire systems\cite{Takesaki-CM-2016,TangJilin-NanoResearch-2023}. Furthermore, this framework is consistent with the experimental realization of nearly $100\%$ AB-stacked BLG single crystals on ultraflat Cu(111)/sapphire substrates\cite{ZhangJincan-NatComm-2023}.

Collectively, temperature tuning and epitaxial strain engineering provide complementary approaches for controlling graphene layer number during CVD growth. Combined with the strategies proposed in our previous work\cite{Wangtt-PRM-2025}, the present framework establishes a versatile parameter space for synthesizing high-quality bilayer graphene.

\section{Conclusion}
In summary, we establish a generalized kinetic framework for graphene CVD growth on Cu substrates by incorporating strain-modulated surface kinetics and chemical desorption into the phase-diagram description developed in our previous work.
Through first-principles calculations and rate-equation analysis, we show that the complex CVD growth process is governed by the competitive balance between first-layer lateral expansion and second-layer nucleation, as modulated by the coupled kinetic parameters $\alpha$, $\Gamma$, and $Z$. 
Our results show that tensile strain fundamentally modifies this competition. While the $i^*=1$ regime remains only weakly affected, tensile strain substantially expands the BLG growth window for $i^*>1$ through the distinct strain responses of first- and second-layer nucleation energetics. In contrast, chemical desorption introduces an additional kinetic suppression mechanism that inhibits BLG formation in the high-$\Gamma$ regime through the desorption parameter $Z$. These findings establish a generalized phase diagram directly connected to experimental parameters, including temperature, hydrogen partial pressure, and epitaxial strain. Beyond explaining experimentally observed trends in graphene CVD, the present framework provides a predictive and experimentally navigable parameter space for the rational synthesis of high-quality bilayer graphene and other two-dimensional materials.

\begin{acknowledgments}
	The numerical calculations were carried out at the High	Performance Computing Center of Nanjing University. This
	work was supported by the National Natural Science Foundation of China (Grants No. 12188101, No. 12274211) and Fundamental and Interdisciplinary Disciplines Breakthrough Plan of the Ministry of Education of China (Grant No. JYB2025XDXM409).
\end{acknowledgments}
\bibliography{reference.bib}
\end{document}